\documentstyle[prb,aps,amsfonts,amssymb,epsfig,floats]{revtex}
\begin{document}
\draft
%\preprint{}
%
%following two lines are for two column format
\twocolumn[\hsize\textwidth\columnwidth\hsize\csname
@twocolumnfalse\endcsname
\title{Role of electron-phonon interaction in the strongly correlated
cuprates superconductors}
\author{Z. -X. Shen$^1$, A. Lanzara$^{1,2}$, S. Ishihara$^3$, and N.
Nagaosa$^{3,4}$}
\address{$^1$Department of Physics, Applied Physics and Stanford Synchrotron
Radiation Lab., Stanford University, CA 94305}
\address{$^2$Advanced Light Source, Lawrence Berkeley National Lab., Berkeley, CA
94720}
\address{$^3$Department of Applied Physics, University of Tokyo,Bunkyo-ku, Tokyo 113
-8656, Japan}
\address{$^4$Correlated Electron Research Center, Tsukuba 305-0046,  Japan}
\date{\today}
\maketitle

\begin{abstract}

Using high resolution angle-resolved photoemission data in
conjunction with that from neutron and other probes, we show that
electron-phonon (el-ph) coupling is strong in cuprates
superconductors and it plays an important role in pairing. In
addition to the strong electron correlation, the inclusion of
phonons provides a theoretical framework explaining many important
phenomena that cannot be understood by a strongly correlated
electronic model alone. Especially it is indispensable to explain
the difference among materials. The phonons with the wave number
around the $(0,q_x)$ and $(q_x,0)$ axes create the d-wave pairing
while that near $(\pi,\pi)$ are pair breaking. Therefore the
half-breathing mode of the oxygen motions helps d-wave
superconductivity.
\end{abstract}

\vskip2pc ]

It has been a long-standing question whether a strongly correlated
electronic model of the CuO$_2$ plane, such as the t-J or Hubbard
model alone, can explain the essential experimental observation of
superconductivity in cuprates oxides. On the one hand, such a
model has been remarkably successful in explaining many important
physical properties, most notably the property of the undoped
insulator and the renormalization of charge dynamics in it, by
spin dynamics from t to J scale\cite{AndersonRVB,ShenSawatzky} and
in predicting a spin gap\cite{Kotliar,Affleck,Fukuyama}. On the
other hand, important questions have been raised. The first is the
observation that, while the CuO$_2$ plane conductivity is
essentially the same for various families of cuprates, their T$_c$
vary by at least an order of magnitude\cite{Battlog,Andersonbook}.
The second is the observation that the phonons and lattice effects
are clearly present in these materials\cite{bookphonons},
following the original assumption that the Jahn-Teller (JT)
polarons might be important for
superconductivity\cite{MullereBernodz}. There is currently no
consensus on the above issues.

Within the context of cuprates superconductors, it is difficult to
understand some of the material specific properties by considering
only the electronic degree of freedom. Fig. 1a summarizes the
systematic in the superconducting gap size ($\Delta$), together
with the transition temperature (T$_c$) shown in Fig. 1c. Unlike
T$_c$, which can be depressed by phase fluctuations in the
underdoped regime \cite{Doniach,Uemura,KivelsoneEmery}, the
superconducting gap essentially reflects the pairing strength. It
can be clearly seen that the pairing strength for the p-type
cuprates is very strong, with gap sizes that are at least an order
of magnitude larger than conventional superconductors. Further,
the maximum T$_c$ of each family is controlled by the maximum
superconducting gap size. This is most dramatically illustrated by
the HgBa$_2$Ca$_{n-1}$Cu$_n$O$_{2(n+1)}$ (Hg1223 for n=3)
compound, which has a much larger gap size as well as T$_c$. Fig
1a also clearly shows a discrepancy between p- and n- type
materials, as superconductivity in a n-type material appears to be
very fragile with a much weaker pairing strength. The
electron-hole asymmetry is particularly perplexing in the context
of spin pairing only, as magnetism is very strong and persists to
a much wider range in the n-type
superconductor\cite{ntypediagram}. It is natural to ask whether
one can gain more insight into this issue by considering the
lattice degree of freedom.

Considering cuprates superconductors in a larger context, there
are other reasons to investigate the role of phonon as exemplified
by the following questions. Why the lattice effect is so important
to the properties of manganites\cite{manganitereview} and
nickelates \cite{nichelate}, but completely unimportant to
cuprates? Why all the high-temperature superconductors with
T$_{c}$ larger than 30K, MgB$_{2}$\cite{Akimitsu}, doped C$_{60}$
\cite{hebard,BattlogC60}, BKBO\cite{Cava} and cuprates, contains
light elements of boron, carbon and oxygen and have strong bonds?
The standard theory of electron-phonon interaction does not give
high-T$_{c}$ and d-wave pairing, but the issue of electron-phonon
coupling\cite{Kulicreview} was not scrutinized to the same extend
as the strongly correlated theoretical models\cite{reviewScience}.

This paper addresses these problems by providing experimental
evidence pointing toward phonon to be also an essential player:
(i) the phonon strongly influences the electronic dynamics, (ii)
the strength of the el-ph coupling correlates with that of
pairing. These finding suggests that both electron-electron
(el-el) and el-ph interactions are essential ingredients for
superconductivity in cuprates. Especially phonon effect is
indispensable to explain the difference among materials showing
different superconducting T$_c$. There are two types of the el-ph
interaction, i.e., the diagonal and off-diagonal one. The former
is the coupling of the atomic (oxygen) displacement to the
charge at the copper site, while the other is the modulation of
the transfer integral due to the phonons. Deriving the effective
t-J model from the three-band d-p model according to Zhang-Rice
\cite{zr}, we obtain both diagonal and off-diagonal interactions
whose ratio are different between p- and n-type cuprates. The
off-diagonal el-ph interaction creates the d-wave pairing while
the diagonal one tends to suppress it. Therefore the weaker d-wave
superconductivity in n-type cuprates, where the diagonal el-ph
interaction is dominant, is consistent with the less phonon
contribution to the pairing. We also discuss the issue of vertex
correction for the electron-phonon coupling.

\begin{figure}[t!]
\centerline{\epsfig{file=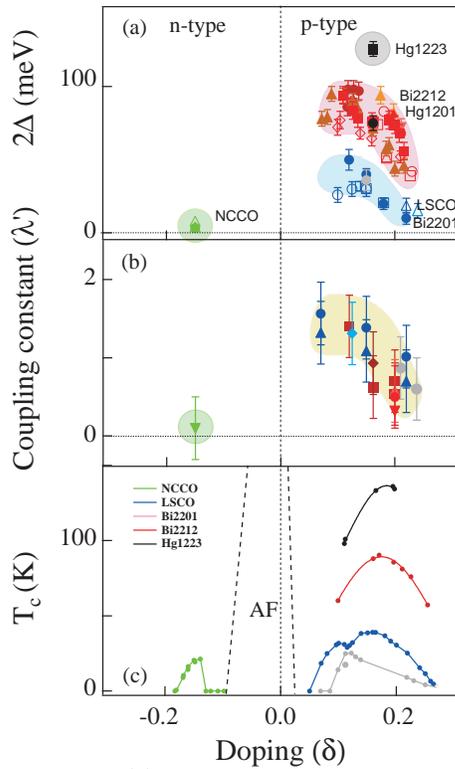,width=6cm,clip=}} \caption{In
panel (a) we plot the superconducting gap determined by
photoemission spectroscopy (full symbol) \cite{photo} and
tunneling spectroscopy (empty symbols) \cite{tunn} as reported in
the literature as well as determined from our data. In panel (b)
the doping dependence of the coupling constant is reported for
LSCO, Bi2201 and Bi2212 (right side) and NCCO (left side) as
determined by angle resolved photoemission, along the nodal
direction ($\Gamma-Y$). In panel (c) (right side), we report the
doping dependence of the critical temperature for four different
p-type systems: Hg1223, Bi2212, Bi2201 and LSCO as reported in the
literature \cite{tc}. On the left side of the same panel, we
report the critical temperature for the NCCO \cite{tc}. The shaded
area are a guide to the eyes.}
\end{figure}

In a strong coupling conventional superconductor the  superconducting transition
temperature is
\begin{equation}
T_c = { {\omega_0} \over { 1.45 } } \exp \biggl[ { { -1.04 ( 1 + \lambda ) }
\over { \lambda - \mu^* (1+0.62 \lambda) } } \biggr]
\end{equation}
where the notations are standard\cite{sca}. For $T_c$ to increase,
one in principle should increase the phonon frequency $\omega_0$.
This, unfortunately, leads to two problems. The first is the
Coulomb repulsion $\mu^* = \mu/( 1 + \mu \ln(E_F/\omega_0) )$,
which increases rapidly with increasing  $\omega_0$, since the
retardation effect is less effective. The second is that higher
frequency means smaller $\lambda$, because $\lambda = C/M
\omega_0^2$, where $C$ is a constant for a given class of
materials.  In order to have a large $\lambda$, as well as high
$\omega_0$, one has to increase the constant $C$, which often
leads to structural phase transitions. These two factors conspire
to limit the value of T$_c$ in conventional superconductors. There
are several band calculations including the el-ph interactions
\cite{freeman,weber,andersen}. The values shown in Fig. 1 for
cuprates make it clear that the conventional wisdom with phonons
will not be able to generate the observed gap size, and thus T$_c$
\cite{weber}.

It is evident that eq.(1) can not be applied directly to cuprates,
where the strong el-el interaction plays essential role. However
we will use this equation below for a guide of the discussion,
which is not unreasonable because the qualitative dependence of
the transition temperature on the coupling constant $\lambda$,
phonon frequency $\omega_0$, and the Coulomb interaction $\mu^*$
remains to be correct.

Although the conventional el-ph interaction alone can not explain
the high-Tc, recent high-resolution angle-resolved photoemission
(ARPES) data suggest that phonons may nevertheless be an essential
player. By analyzing photoemission data from
Bi$_{2}$Sr$_{2}$CaCuO$_8$ (Bi2212), Bi$_{2}$Sr$_{2}$CuO$_6$
(Bi2201) and La$_{2-x}$Sr$_{x}$CuO$_4$ (LSCO), in addition to
earlier data from Bi2212 \cite{Valla,Bogdanov,Kaminski}, we
concluded that the quasiparticles in p-type cuprates are strongly
coupled to phonons in the frequency range 50-80meV\cite{Lanzara},
in contrast to that of the n-type material\cite{peter2}. We see a
dramatic change in the "quasiparticle" velocity in the form of a
break in the dispersion near this energy scale, as shown in Fig. 2
(a-c) where the dispersion along the nodal direction for three
families of p-type cuprates are plotted. This result is very
robust, and the details of the fit can be found
elsewhere\cite{Bogdanov,Lanzara}. In related data, a drop in the
quasiparticle scattering rate below this energy scale is
observed\cite{Bogdanov,Kaminski}, which is consistent with optics
data reporting a rapid drop of the scattering rate, 1/$\tau$, at
similar energy\cite{timuskreview}. The observed behavior, a sudden
change in the dispersion and a drop in the scattering rate, is
very reminiscent of quasiparticle  coupled to a sharp collective
mode. The famous 41 meV neutron mode and phonons are the only two
sharp collective modes we are aware of. Since this effect is seen
in all compounds and is not associated with the superconducting
phase, as the phenomena are seen well above T$_c$, we rule out the
neutron mode possibility \cite{Norman}, as neutron modes are only
seen in YBa$_2$Cu$_3$O$_{7-x}$ (YBCO) and Bi2212 and are only
associated with superconductivity. We note here that the overdoped
Bi2201 data was collected at 30K, six times higher than T$_c$.
Alternatively one may try to explain the data by the opening of a
gap elsewhere on the Fermi surface. However, the superconducting
and pseudogap in LSCO, Bi2201 and Bi2212 are very different, but
the "kink" energy is very similar, so this explanation does not
work. Thus we are left with phonons as the only surviving
candidate to explain our data\cite{Lanzara}.

The phonon interpretation receives additional support from
comparison between photoemission and neutron data. Neutron
scattering data from LSCO and YBCO suggest that the zone boundary
phonon couples most strongly to doped charge\cite{Egami} and
softens significantly with doping. In the case of LSCO, where a
direct comparison between neutron and photoemission is possible,
the phonon mode appears at 70meV. As shown by the thick arrow in
Fig. 2c, the phonon energy coincides with the "kink" energy in the
dispersion, providing a very strong piece of direct evidence for
the phonon being the mode responsible for the effect seen in
quasiparticle dynamics. In a separate analysis we show the
striking similarity between photoemission data from Bi2212 with
that of the Be(0001) surface state where the el-ph coupling is
known to be strong \cite{Lanzara}. In both cases, a peak-dip-hump
structure is clearly seen in the spectra, as expected in theories
for electrons coupled to a collective mode \cite{sca,NormanDing}.
This analogy again provides strong support for the phonon
interpretation. In summary, we believe that this ensemble of data
makes a compelling case for a strong coupling between
quasiparticles and phonons.

In conventional theory the phonon correction to the electronic
velocity is given by $v(k)=v_0/(1+\lambda)$, where $v_0$ is the
bare electron velocity (below the phonon frequency $\omega_0$),
$v(k)$ is the dressed velocity (above the phonon frequency) and
$\lambda$ is the el-ph coupling constant\cite{Mermin}. In a real
experiment, one cannot obtain $\lambda$ without knowing the bare
velocity that can be different from band structure value due to
el-el interaction. We have extracted a similar quantity which we
call $\lambda$' using the velocity obtained from the experimental
dispersion above the phonon frequency to approximate the bare
velocity. In this case, we define the velocity ratio below and
above the phonon energy as 1/(1+$\lambda$'). Because the high
energy velocity is an overestimate of the bare velocity,
$\lambda$' is an overestimate of $\lambda$. Within a specific
model such as Debye model, one can see that this overestimate can
be 20-30$\%$. We use the experimental quantity $\lambda$' for this
discussion as the systematics are the same. We find that the
velocity ratio of p-type materials is close to two near optimal
doping and increases with underdoping. While the quantitative
value is model and procedure dependent, the data suggest that
el-ph coupling is strong in these materials. The el-ph coupling
constant ($\lambda$') as a function of doping is plotted in Fig.
1b. We note that the frequency scale of the phonon here is an
order of magnitude higher than that of conventional
superconductors. In principle, given the high $\lambda$' value, it
can deliver a pairing strength that is an order of magnitude
higher with respect to the conventional superconductors, if one
can figure out a way to deal with the $\mu^*$ problem. As we will
elaborate later, the d-wave pairing helps this issue.

In contrast to p-type materials, the kink effect is much weaker
and basically not discernible along the nodal direction for n-type
materials, Fig. 2d\cite{peter}, indicating a much weaker el-ph
coupling. This significantly weakened effect is consistent with
optical data, where the drop in 1/$\tau$ is found to be much
smaller in n-type material\cite{Basov,Calvani}. Intuitively this
is not surprising, as these high energy phonons must involve
in-plane oxygen atoms, as suggested by different
experiments\cite{stripe}. For p-type materials the doped carriers
predominately go to the oxygen site, while they go to the copper
site (upper Hubbard band) in n-type material. This qualitative
difference is expected to give a large difference of the el-ph
interaction between n- and p-type cuprates as discussed below.

A glance to Fig.1a and Fig.1b makes it clear that there is a
correlation, as a function of doping, between the pairing gap size
and the el-ph coupling constant. To first order, the most striking
feature in both Fig.1a and Fig.1b is that the electron doped and
hole doped materials are very different. The much weaker el-ph
coupling strength explains smaller pairing gap in electron-doped
material. Within a single band t-J or Hubbard model with only
nearest neighbor hopping integral, one would expect particle-hole
symmetry. The other clear correlation is that the pairing strength
of p-type material decreases with doping in Fig.1a, as does the
el-ph coupling constant in Fig.1b. We  believe that the
correlation seen in these figures makes a strong case for the
el-ph coupling being a key to pairing.

\begin{figure}[t]
\centerline{\epsfig{figure=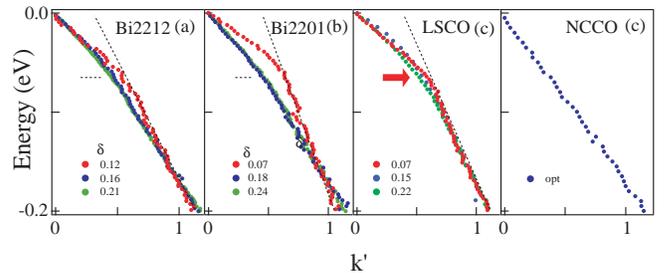,width=9cm,clip=}}
\vspace{.2cm} \caption{The quasiparticle dispersions vs the
rescaled momentum (k') are reported for three p-type materials
systems; Bi2212 (panel a)\cite{Bogdanov,Lanzara}, Bi2201 (panel
b)\cite{Lanzara} and LSCO (panel c)\cite{Lanzara}. The arrow
indicates the frequency value obtained by inelastic neutron
diffraction data. The dispersions are compared with the n-type
superconductor NCCO (panel d)\cite{peter2} along $\Gamma$Y. The
rescaled momentum, k', is defined as ($k-k_F$)/$k_{170meV}$. The
dotted lines are guides to the eye and are obtained fitting the
high energy part with a linear function.}
\end{figure}

As for the systematic on the gap size in different p-type
materials, the situation is more complex with the following
factors being potentially important. It is known that the material
dependent copper and apical oxygen distance correlates with T$_c$
at optimal doping, T$_c^{max}$. It has been discussed that this is
related to the local stability of the Zhang-Rice singlet and its
effective transfer integral \cite{taniguchimaekawa}. Another
electronic structure analysis reduces this down to a correlation
between T$_c^{max}$ and the range of the intralayer
hopping\cite{Pavarini}. Hg1201 is found to have not only a larger
intralayer hopping than LSCO/Bi2201,  but also a larger interlayer
hopping. The latter is due to the on-top stacking of the CuO$_2$
layers in Hg1201. In LSCO/Bi2201 the interlayer hopping is found
to be substantially smaller due to the body-centered tetragonal
stacking.  Difference in hopping may be a reason \cite{andersen}
for $\Delta$ in Bi2212/Hg1201 (T$_c^{max}$=90K) and LSCO/Bi2201
(T$_c^{max}$=30-40K) to segregate into two groups. For the much
higher $\Delta_{max}$ in Hg1223 there is the additional
possibility of a resonance between the phonon and the
superconducting gap, being the two energies close to each other.
Optics experiment saw in fact a drop in 1/$\tau$ in Hg1223 system
at a frequency 30$\%$ higher than the one observed in Bi2212
\cite{Timusk}, which scales exactly with the pairing gap (and
T$_c$) difference. The reason for the difference in Hg1223 is
maybe the reduced strain on the CuO$_2$ sheets by other layers in
this material, a factor related to the $\mu*$ problem as we will
discuss later.

Now we turn to the more detailed description of the theoretical
consideration. Fig.3a shows the displacement pattern of the zone
boundary half-breathing mode (q=($\pi,0)$), which corresponds to
the important phonon with frequencies 50-80meV. There are several
ways to consider its interaction with the electronic system. The
usual approach is to consider the diagonal el-ph interaction in
the particle number, namely the oxygen displacement is coupled to
the electron density $n_i$ in the i-th Cu
orbital\cite{Bulut,song}.  When one considers the dispersionless
phonons with the momentum-independent coupling constant, this
in-plane oxygen displacement is pair-breaking for d-wave because
it gives rise to the repulsive interaction between the electrons
on the nearest neighbor Cu orbitals. This is more appropriate when
the carriers are doped into the Cu orbital as in n-type cuprates,
but not when they are doped into the oxygen orbital as in p-type
cuprates. With the strong el-el interaction, the vertex correction
to the el-ph interaction becomes
crucial\cite{zeyher,kim,castellanivertex}. Especially this
diagonal el-ph coupling is irrelevant in the undoped case because
the charge fluctuation is completely suppressed there. In this
case the only relevant el-ph interaction is in the spin sector,
i.e., the modulation of the exchange interaction $J$ due to the
phonon, which is related to  the modulation of $t_{ij}$ described
below \cite{sawatzky}. With the small doping concentration $x$,
the diagonal el-ph interaction is still not so effective because
the vertex correction gives the reduction factor $x$ to the
effective coupling constant\cite{zeyher,kim,castellanivertex}.
Further, the absolute value of the dielectric constant
$\epsilon(\omega)$ at the energy of the gap $\omega=2\Delta$ is
much larger than one for the superconducting materials, and the
diagonal el-ph interaction is expected to be screened and reduced
considerably.

Therefore we will focus below on the off-diagonal el-ph
interaction. For the holes, doped predominantly into the oxygen
orbitals, the direct el-ph coupling originating from the shift of
the $d$-level due to the oxygen displacement is missing. In this
case, the el-ph interaction should be derived in the framework of
the effective t-J model where the strong correlation and the
exchange interaction are taken into account. A realistic model of
electronic structure is the d-p model, which considers the d- and
p- orbitals as well as the strong Coulomb interactions. When we
reduce the d-p model  into the effective single-band t-J model,
the transfer integral  $t_{ij}$ between the Zhang-Rice
(ZR)\cite{zr} singlet states, at  $i$ and $j$ sites, is given by
the second order process in the  hybridization $t_{dp} \sim 0.8
eV$ between the d- and p-orbitals as
\begin{equation}
t_{ij} \cong { {t_{dp}^2} \over {\Delta_{dp} } }
\end{equation}
where $\Delta_{dp}$ is the energy difference between the p- and
d-levels \cite{zr,maekawa}. When the negatively charged oxygen
ions approach (go away from ) the copper, the d-level energy
increases (decreases). This reflects the fact that the oxygen
displacement modulates the energy level of d-orbital as
\begin{equation}
\delta \varepsilon_d = g u(i)
\end{equation}
where $u(i)$ is  the  linear combination of the oxygen  displacements surrounding
the copper $i$ given by
\begin{equation}
u(i) = u_x(i-x/2) - u_x(i+x/2) + u_y(i-y/2) - u_y(i+y/2),
\end{equation}
where $u_\mu(i \pm \mu/2)$ ($\mu = x,y$) is the displacement along
the $\mu$-axis of the oxygen between the two coppers at $i$ and $i
\pm \mu$. On the other hand the energy level of the oxygen
p-orbitals does not change in linear order in $u$'s. Therefore the
charge transfer energy $\Delta_{dp} = \varepsilon_p -
\varepsilon_p$  in the hole picture is given by  $\Delta_{dp} =
\Delta_0 - g u$, which leads to the off-diagonal el-ph interaction
through eq.(2). $\Delta_0$ is about 2eV, while $g$ and $u$ are the
el-ph coupling constant and the displacement, respectively. Then
the modulation of the transfer integrals $t_{ij}$ is given by
\begin{equation}
\delta t_{ij} \cong -g { {t_{dp}^2} \over {\Delta_{0}^2 } } (
u(i) + u(j)).
\end{equation}
This is the off-diagonal coupling \cite{sawatzky}. It is noted
here that $\Delta_{dp}(i)$ and $\Delta_{dp}(j)$ ( $u(i)$ and $u(j)$ )
appear symmetrically in  $t_{ij}$.  On the other hand,  the
modualtions of $t_{ij}$ due to those of $t_{dp}$`s  cancel in
linear order in $u$'s. In Fig. 3a, we take the other view focusing
on the displacement of one oxygen, and see the modulation of the
transfer integrals $t_{ij}$. Taking other sites into account the
modulation in t$_{ij}$, parallel to the displacement (green
colored), cancel out and only the t$_{ij}$ perpendicular to the
displacements are modulated.

It is noted here that the transfer integrals $t_{pp}$'s between
the oxygen p-orbitals are also relevant to $t_{ij}$ in the t-J
model, and the modulation of $t_{pp}$'s due to the oxygen
displacements gives another channel for the off-diagonal el-ph
interaction. We believe this contribution is the main reason why
the off-diagonal el-ph interaction is different between the p- and
n-type materials. The wave function of ZR-singlet has more weight
on the oxygen p-orbitals in the p-type case and is expected to
have larger el-ph interaction due to this mechanism. It should be
noted here that the el-ph interaction corresponding to the
modulation of the exchange interaction is also present, and is
expected to be dominant in the underdoped region \cite{sawatzky}.
The exchange interaction can be represented as $J {\vec S}_i \cdot
{\vec S}_j = - J \chi_{ij}^\dagger  \chi_{ij} + {\rm constant}$ in
terms of the bond variable $\chi_{ij} = \sum_{\sigma} C^\dagger_{i
\sigma} C_{j \sigma}$. By replacing $ - J \chi_{ij}^\dagger
\chi_{ij} \to  - J (<\chi_{ij}^\dagger>  \chi_{ij} +
\chi_{ij}^\dagger  <\chi_{ij}>)$ in the mean field approximation,
this interaction can be reduced to the same interaction as eq.(5).
An important feature of this exchange modulation el-ph interaction
is that it is free from the reduction due to the vertex correction
by the factor of $x$ as shown below in eq.(6). It represents the
fact that it survives in the half-filled case. In Table 1 are
summarized the el-ph interactions in p-type and n-type materials.

%\vskip2pc]
%
\vskip 0.5cm
\begin{center}
\begin{tabular}{l|l|l}
&
diagonal el-ph &
off-diagonal el-ph \\
\hline
p-type &
weak and reduced by $x$ &
strong
 \\
\hline
n-type&
strongest but is reduce by $x$ &
weak
\\
\hline
\end{tabular}
\vskip 0.3cm
Table 1. Classification of electron-phonon interactions.
\end{center}
\vskip 0.5cm

Below we focus on the off-diagonal el-ph interaction, which is most relevant in the
underdoped region due to the vertex corrections and  can be represented by the
standard el-ph Hamiltonian as
\begin{equation}
H_{\rm el-ph} =  { 1 \over \sqrt{N} } \sum_{k,q,\sigma}  g(k,q) u_q C^\dagger_{k+q
\sigma} C_{k \sigma} .
\end{equation}
where $g(k,q)$ has the characteristic $k,q$-dependence for
off-diagonal interactions as $g(k,q) \propto (\sin(q_x/2) +
\sin(q_y/2)) \times ( \cos(k_x+q_x) + \cos(k_x) + \cos(k_y+q_y) +
\cos(k_y) )$. These momentum dependence is essential for the
pair-creating of d-wave superconductivity as shown below.

The above off-diagonal el-ph interaction, which is tied to the
change of bond (rather than site) by lattice vibration, is
significantly enhanced by the strong Coulomb interaction on the Cu
site\cite{ishihara}. This is expressed by the vertex correction
for the el-ph interaction, which is represented by the following
formula for the effective el-ph coupling constant \cite{nagaosa}.
\begin{equation}
g_{\rm eff.}(k,q)  = \gamma(q) g(k,q) = { {g(k,q)} \over { 1 - J \Pi_{\chi} (q) } }
\end{equation}
where $\Pi_{\chi}(q)$ is the generalized susceptibility for bond
variable $\chi_{ij} = \sum_{\sigma} C^\dagger_{i \sigma} C_{j
\sigma}$, and  the factor $\gamma(q) =  { {1} \over { 1 - J
\Pi_{\chi} (q) } } $ is the usual Stoner enhancement factor in the
RPA approximation. As described below, this interaction creates
d-wave pairing. According to the BCS theory, we integrate over the
phonons to create the effective el-el interaction.
\begin{eqnarray}
H_{\rm eff} &=& -{ 1 \over N } \sum_{k,k',q,\sigma,\sigma'} g_{\rm eff.}(k, q) g_{\rm
eff.}(k',-q) <u_q u_{-q}>
\nonumber \\
&\times& C^\dagger_{k+q \sigma} C_{k \sigma} C^\dagger_{k'- q \sigma'}
C_{k' \sigma'}
.
\end{eqnarray}
Here we have assumed for simplicity  the limit of the high phonon
frequency, which is justified marginally as follows. In the present case,
$\omega_0 \cong 50- 80 meV$ is comparable to $ J \cong 140 meV$
which determines the effective band-width. We have found that the
qualitative conclusion does not change when we consider the case
of more realistic bandwidth determined experimentally.
\begin{figure}
\centerline{\epsfig{file=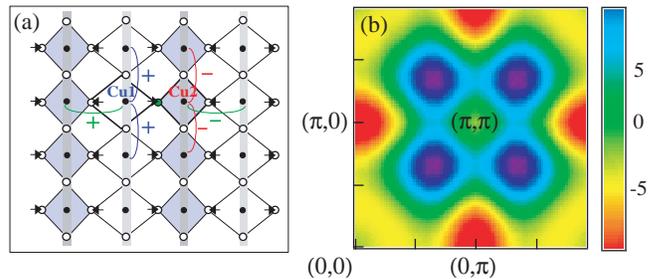,width=8.5cm,clip=}}
 \vspace{.2 cm}\caption{In panel a) we report the displacement pattern
of zone-boundary half-breathing mode of the CuO$_2$ planes. The
black dots represent the Cu atoms, while the white ones the oxygen
atoms. The off-diagonal el-ph interaction are shown. The
displacement of O at the center modulates the energy of d-orbitals
at the two Cu sites, Cu1 and Cu2. The effective transfer integral
t$_{ij}$ perpendicular to the displacements are modulated
\cite{Egami}. In panel b) we report the contour plot of the
momentum dependence of the pairing $e(q)$. Negative $e(q)$
indicates pair-creating region, while positive $e(q)$ indicates
pair-breaking region. The pairing force  is zero (green) around
the nodal  region while it decrease gradually, reaching  a minimum
(red) approaching the ($\pi,0$) region. }
\end{figure}

In the case of s-wave pairing, the Debye cut-off is critical and
only below it the effective attractive force is effective.
Therefore the smaller frequency reduce the pairing force and
enhances the pair-breaking. For the d-wave pairing as will be
assumed below, on the other hand, the momentum-dependence is more
important, and the sign of the interaction is not relevant. It is
because the constant part of the effective interaction does not
contribute due to $\sum_k < C_{k \uparrow} C_{-k \downarrow} > =
0$. We have also checked that the retardation effect does not
change the conclusion qualitatively. Our argument is based on the
fact that the strong correlation and exchange interaction leads to
d-wave superconductivity. We calculated the energy in the BCS
approximation for d-wave superconductivity, with the order
parameter $\Delta_k = < C_{k \uparrow} C_{-k \downarrow} > = (\cos
k_x - \cos k_y)/\sqrt{ \cos^2 k_x + \cos^2 k_y } $ appropriate for
the half-filled case. Then we obtain
\begin{eqnarray}
<H_{\rm eff}>_{\rm BCS} &=& -{ 2 \over N } \sum_{k,q} |g(k, q)|^2 |\gamma(q)|^2 <u_q
u_{-q}>
\Delta_k \Delta_{k + q}
\nonumber \\
&=& \sum_q  <u_q u_{-q}> e(q) |\gamma(q)|^2
\end{eqnarray}
where we have done only the $k$-integral in the last line of the
above equation. It is noted here that $<u_q u_{-q}> \propto
\omega_q^{-2}$ and hence the experimental information can be put
through this factor. Therefore $e(q)$ is weighted more in the
momentum region where the strong el-ph coupling is observed and
the dispersion shows softening, i.e., along the axis $(q_x,0)$ and
$(0,q_y)$. In Fig. 3(b) we show the contour plot of $e(q)$
\cite{nagaosa}. It is noted here that the oxygen displacement
contributes strongest to the d-wave pairing in exactly the region
of q-space, namely along the line $q =(q_x,0)$ and $q=(0,q_y)$,
where the strong coupling to the electrons is observed
experimentally \cite{Egami,Braden}. It should be noted here that
$u(i)$ is given by the linear combination of the various phonon
modes other than the half- breathing mode in general, and also
even the displacements of the apical oxygen play similar roles to
give off-diagonal el-ph interaction. On the other hand the phonons
with the momentum around $q=(\pi,\pi)$ are decoupled from the
electrons due to the momentum dependence of the off-diagonal el-ph
coupling constant $g(k,q)$ given below eq.(5). In this momentum
region, the pair-breaking effect originates in the case of
diagonal el-ph coupling. In experiments, the anomalous dispersion
of this half-breathing mode has been observed and the zone
boundary region modes show the softening and couple to the
electrons  more strongly  than  the zone center region modes
\cite{Egami}. In another paper\cite{Braden} it is found that el-ph
coupling is strong from (0.5$\pi$,0) to ($\pi$,0). In any case the
finite wave number phonons play important role \cite{mihailovic}.
Integration of this range gives pair creating results.  Especially
it is found that $|\gamma(q)|^2$ strongly depends on $q$, and
enhances the coupling around $(0,\pi)$ and $(\pi,0)$ points. This
corresponds to the tendency towards the spin-Peierls (dimer) state
or stripe formation, and favors the d-wave pairing because the
pair-creating momentum region is enhanced by this vertex
correction.

Now we comment on the n-type cuprates. Although our theory can not
conclude definitely that el-ph interaction is weaker in n-type,
the smaller kink in Fig. 2 and the smaller scattering rate
$1/\tau(\omega)$ in optical experiment \cite{Basov} strongly
suggests it. In particular, the el-ph coupling in n-type cuprates
is much weaker along the $(\pi,\pi)$-direction (although el-ph
coupling, or the kink effect is seen along other directions), this
is consistent with the weaker off-diagonal el-ph interaction in
the n-type material. Also the tunneling data from NCCO has been
interpreted by the phonons with lower frequencies, which gives
lower $T_c$. Therefore we can not expect much from the el-ph
interaction for the pairing force.  It is not clear if this
reduced pairing force due to el- ph interaction is enough to
explain the $T_c$ in n-type material. It is possible that the
antiferromagnetic fluctuations contribute to the d-wave pairing as
discussed extensively \cite{Scalapinoreview,Pinesreview}.
Therefore certain phonon channels can be compatible with d-wave
pairing. Here we note that our identification of the phonon
structure in ARPES by the half-breathing $q =(\pi,0)$ phonon is
mostly motivated by the neutron experiment showing it to soft
significantly with doping \cite{Egami}. The energy consideration
of the photoemission data is also consistent with this mode.
However, the experimental uncertainty in the photoemission data
analysis also allows the possibility of the lower energy phonons
in the range of 40-50 meV. Part of this is related to the
uncertainty of the superconducting gap size which modifies the
kink position. Therefore the half-breathing mode in our discussion
should be regarded as a linear combination of various modes. This
is particularly true for the in-plane buckling mode (40-50 meV).
According to a band calculation \cite{andersen96}, both the
half-breathing mode and this buckling mode show d-wave pair
creating tendency near $q \sim (0.5 \pi - \pi,0)$ and pair
breaking near $q \sim (\pi,\pi)$. Raman experiments have shown
very strong softening of this phonon through T$_c$\cite{Cardona}.
We hope that our discussion stimulates more theoretical
investigations on the el-ph interaction in strongly correlated
electron systems, as the data in Fig. 1 and Fig. 2 strongly
suggest phonon to be a key player for superconductivity,
independent of specific theory. It is probably true that many
phonons need to be considered.

Fig. 4 proposes a phase diagram that contains essential
ingredients for the key physics. At very high energy and
temperature, there is an energy scale (or cross-over line) T$_0$,
which is related with J scale physics caused by strong Coulomb
interactions. At lower energy scale (or temperature) phonons helps
pairing which is allowed with compatible magnetism to deliver the
pairing, yielding the intermediate pairing energy scale
T$_{phonon}$ (T$_{ph}$). This two scales energy scheme naturally
explains the observation that there are two kinds of pseudogaps in
underdoped cuprates, as strongly indicated by ARPES data near
($\pi,0$)\cite{Marshall}. The high energy one (hump) connects
smoothly to that of the insulator\cite{Laughlin}. The low energy
one (leading edge) is related to the superconducting
gap\cite{Marshall}.

A key issue for the classical phonon theory is the problem with
$\mu^*$. Since J has higher energy (Fig. 4), the antiferromagnetic
interaction dictates that the pairing state can only be of d-wave
symmetry. This suppresses the s-wave pairing instabilities of some
phonons while cooperates with the d-wave pairing processes of
other phonons. Since the d-wave state has a node at the origin,
this significantly reduces the $\mu^*$ problem or even changes
sign, because there is no amplitude on the same site, and the
on-site repulsion is not an issue. We note that while $\mu*$ in
eq. (1) is introduced in the context of s-wave superconductor as a
pair-breaker, the $\mu^*$ in the discussion here should be either
positive or negative depending on the model without the el-ph
interaction. Another critical role of the strong Coulomb
interaction is that it significantly enhances the off-diagonal
el-ph coupling near the quasi-resonance of the p- and d- levels of
cuprates\cite{ishihara}. This is understood in the vertex
correction in eq.(7) due to the near instability of the system
towards the spin-Peierls type ordering and/or the bond-centered
stripe formation \cite{Sachdev}. As we have indicated earlier, the
high phonon frequency of 50-80meV and the strong coupling constant
can deliver a pairing strength that is an order of magnitude
higher (and thus high-T$_c$) if the $\mu^*$ problem can be
significantly reduced.

Another factor to differentiate $\Delta_{max}$ among families of
compounds is the presence/absence of the static distortion. Aside
from theory \cite{taniguchimaekawa,Pavarini}, that is based on the
average structural data, this tendency is re-enforced by local
structural data. As shown by local structural probes, the 70meV
zone boundary phonon couples with dynamic local structural
distortions\cite{stripe}. The onset temperatures of the distortion
(T*) are significantly lower in LSCO, compared to Bi2212 and YBCO,
likely related to stripes\cite{elec}. This coupling with more
static structural distortion in LSCO is harmful to the pairing and
superconductivity explaining the lower value of $\Delta_{max}$,
although $\omega_0$ is comparable. For a similar reason we do not
observe any clear change of $\omega_0$ in the
La$_{1.28}$Nd$_{0.6}$Sr$_{0.12}$CuO$_4$ (NdLSCO) system with
respect to optimally doped LSCO\cite{XJ}. The stripes and
distortion in NdLSCO are static, further suppressing pairing. This
may also be a factor for $\Delta_{max}$ to be the largest in
Hg1223 as the CuO$_2$ planes are the flattest there.

\begin{figure}[t]
\centerline{\epsfig{file=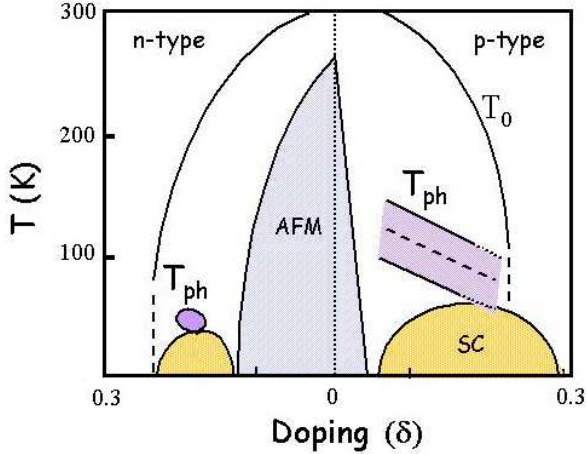,clip=,width=8cm}} \vspace{.1cm}
\caption{A schematic phase diagram for high temperature materials
is presented. The black line represent the T$_0$ line, which is
particle-hole symmetric in the t-J model. The purple shaded area
represents the range where the phonons become important. The
superconducting regime (SC) is indicated by yellow shaded area and
the antiferromagnetic regime (AF) by blue shaded area.}
\end{figure}

The above interpretation receives support from the large change in
the onset temperature T* of the distortion (order of 80K) with
isotope substitution \cite{isotope}. The opposite change in the T*
respect to the one in the T$_c$ (T* increases with increasing the
mass of the isotope), is consistent with the idea that more static
lattice distortions suppress pairing. The local structural
distortion is likely to be similar in Bi2201 and this may explain
why the maximum T$_c$ (30K) for the single layer is much lower
than that of the isostructural single layer of Tl and Hg
compounds. The difference in the dynamics of local structural
distortion between different families and different dopings is
probably related to the mismatch between the CuO$_2$ plane and the
block layers\cite{isotope,locallattice,mercurio}. As one
introduces more CuO$_2$ planes in a unit cell, their structures
are less vulnerable to strains from other layers, and this may
explain why T$_c$ increases with number of layers in Bi, Tl and Hg
based families of cuprates.

We now briefly discuss the connection between the picture which
emerges here and those in the literature. First there are several
authors who considered the interplay between the electron
correlation and el-ph interaction
\cite{ishihara,zeyher,giniyat,yonemitsu,varelog}. One important
conclusion there is that even starting from the momentum
independent el-ph coupling, the vertex function due to el-el
interaction suppresses the large momentum transfer process, and
hence the transport $\lambda_{tr}<<\lambda$ \cite{zeyher}. This is
expected to  resolve the apparent contradiction between the ARPES
and transport data. Namely the el-ph interaction of the order of
$\lambda_{tr} \sim 1$ does not appear in the resistivity while
$\lambda \sim 1$ is needed to explain the ARPES data. It has been
pointed out before that the resistivity data is not incompatible
with the presence of el-ph coupling, contrary to general belief
\cite{philallen}. However these works does not take into account
the modulation of the exchange interaction \cite{sawatzky} which
we believe is the most important for underdoped region.  One clue
is the $x$-dependence of the coupling constant. The usual
treatment in terms of the 1/N expansion and/or the slave boson
method leads to the conclusion that the effective coupling
constant is proportional to $x$ for small $x$, which is not
consistent with the observed $x$-dependence shown in Fig. 1
\cite{kim}. On the other hand, when one consider the off-diagonal
coupling described in eq. (7), the doping will reduce the Stoner
enhancement factor of the bond-order, and hence the coupling
constant is expected to be decreased. We have concentrated on the
zero-temperature superconducting state, and the transport
properties are beyond the scope of this paper. At finite
temperature above $T_c$, the spin-charge separation may play an
important role, and the charge carriers are not the simple
electrons. If the momenta of the carriers are smaller than the
Fermi momentum, the large momentum phonon scattering does not
contribute so much to the resistivity. However this issue is a
subtle one, and we do not go further in detail on it. As for the
coupling to the spin fluctuation as the origin of the kink in the
dispersion and the peak-hump-dip structure, we believe the small
spectral weight of the 41meV-resonant peak invalidates that
scenario \cite{zeyher}.

Several authors looked for the mechanism of superconductivity in
the interplay between the electron correlation and el-ph
interaction \cite{dagotto,kohmoto,varelog}. Here it should be
noted that the recent tunneling experiments give evidence that the
pseudo-gap and superconducting gap may be distinct
\cite{Innov,Kapitulnik,Pan}, while the low energy part of these
gaps seems to be identical. If these picture is correct, which is
still an open question \cite{Fischer}, this pseudo-gap play an
essential roles in the physics of underdoped and optimally doped
cuprates, which can not be captured by the perturbative analysis.
It is natural to consider that the strongest interaction, which is
the Coulomb interaction and/or the spin exchange interaction,
determines the global and high energy physics, while the  phonons
play important roles in the lower energy physics. This is the
picture we take in this paper, assuming for example that the
superconducting gap is d-wave symmetry,  which can not be derived
by el-ph interaction only. Obviously, diagrams similar to Fig. 4
already exist, except that the data in Fig. 1a and Fig. 1b
strongly suggest that the intermediate pairing line is influenced
by phonons.

There are many scenarios for the high energy dynamics and
pseudo-gap formation. One possibility is the antiferromagnetic
short range order based on the Hubbard type models
\cite{yamada,Hubbardreview}.   The second one is the spin singlet
formation based on the RVB picture \cite{AndersonRVB} in terms of
the t-J models\cite{Kotliar,Affleck,Fukuyama,su2}. In these cases,
the role of the phonons has been neglected in most of the
literature although some authors mentioned it implicitly
\cite{kivelsonRVB,Sachdev}. For strong coupling theories there is
no consensus on the lowest energy state in these strongly coupled
models, with flux phase, d-wave pairing, and
anti-ferromagnetically ordered ground states being possible
candidates. In the intermediate energy scales between the
pseudo-gap and superconducting gap, it is expected that the state
can be regarded as the thermal and /or the quantum mixture of the
possible ground states such as the staggered flux  and d-wave
superconducting state in SU(2) gauge theory\cite{su2}. The role of
the el-ph interaction here is to determine the superconducting gap
size and transition temperature, which differ significantly from
material to material. The third scenario is the charge ordering
and/or the stripe formation discussed intensively\cite{CDW}. These
models are related to what we propose, since the stripe approach
also includes two energy scales and furthermore the half-breathing
mode can be regarded as the instantaneous creation of the dynamic
stripe (Fig. 3a). In some theories phonons  are explicitly
considered \cite{neto}, while in others only the electronic
degrees of freedom are stressed\cite{fononi}. For the cases
involving the Hubbard-Holstein model, the discussion mainly
concentrates on the issue of quantum critical points without
specifying the role of these particular phonons\cite{roma}. For
the cases where only the electronic aspects of the stripes are
considered, the two energy scales correspond to the scales when
the stripes and pairs are formed respectively\cite{zachar}. In
this case the pairing of the carriers is purely electronic, driven
by quasi-1D behavior. Independently of the specifics, it would be
interesting to see the role of the electronic stripes on the el-ph
coupling considered in our case. Naturally, charge ordering
couples to lattice distortions. The fact that the el-ph coupling
is strong for phonons with q value from ($0.5\pi,0$) to ($\pi,0$)
could be interpreted as the system tendency to couple with stripes
of 4a to 2a periodicities\cite{Braden}. This issue of coupling to
2a periodicity has been discussed before\cite{Sachdev,neto}. As
shown in Fig. 3a, the ($\pi,0$) phonon has a clear 1D character
with 2a periodicity. This could be regarded as the instantaneous
creation of the dynamic stripes because the  charge should be
accumulated/diluted alternatively along the Cu-O chain
perpendicular to the displacement. Alternatively, one can say that
the systems tendency to have stripes will promote el-ph coupling
with desired q, that is pair creating for d-wave state. Therefore
the stripe theory could be regarded as the strong coupling limit
of the el-ph interaction.

Before leaving the subject of theory, we also note that there are
extensive work on phonons only, without explicit consideration of
the strong Coulomb correlation, notably the polaronic effect and
bipolaron condensation\cite{polarons}. Here we think that the
d-wave aspects are important because of the $\mu^*$ problem in
conventional el-ph coupling theory. Whatever the specific
relationship between our finding and the interesting theoretical
ideas discussed above may be, the data hint strongly that the
el-ph interaction must be considered explicitly. At the same time,
we also stress that the el-ph interaction must be considered in
the context of strong electronic correlation, as the t-J model has
been remarkably successful to describe the electronic structure of
cuprates down to J scale \cite{ShenSawatzky,Hubbardreview}. In
this sense the strongly correlated t-J model does provide a basis
for cuprates physics, with phonons add to it in delivering d-wave
superconductivity with very high-T$_c$.

In summary, we have shown a comparison of photoemission
experiments in conjunction with neutron, and other probes,
providing direct evidence for strong el-ph coupling being
important for pairing. The inclusion of el-ph interaction explains
many experimental observations that cannot be understood by
strongly interacting models of CuO$_2$ planes alone.

We would like to acknowledge P. V. Bogdanov, X. J. Zhou and S. A.
Keller for experimental help. We would like to thank P. Allen, A.
Bianconi, D. Bonn, S. Doniach, T. Egami, T. H. Geballe, S. A.
Kivelson, R. B. Laughlin, D. H. Lee, K. A. Muller, D. Mihailovic,
D. J. Scalapino and G. A. Sawatzky for useful discussion. The
SSRl's work was supported by the DOE, Office of Basic Energy
Science, Division of Materials Science. The work at ALS was
supported by the Office of Division of Materials Science with
contract DE-AC0376SF00098. One of us A.L. would like to thank the
Instituto Nazionale Fisica della Materia (INFM). This work is
supported by the Grant in Aid form Ministry of Education, Culture,
Sports, Science and Technology of Japan.


\begin{references}
\bibitem{peter} The kink effect is basically none existing along
the nodal direction, and the effect is stronger in other
directions.
\end{references}
\end{document}